\documentclass[aps,prl,amssymb,twocolumn,showpacs]{revtex4}

\usepackage{graphicx}

\newcommand{\bea}{\begin{eqnarray}}

\newcommand{\eea}{\end{eqnarray}}

\newcommand{\beq}{\begin{equation}}

\newcommand{\eeq}{\end{equation}}

\newcommand{\lav}{\langle}

\newcommand{\rav}{\rangle}

\newlength{\textwidthm}

\begin{document}

\title{A supercritical superfluid and vortex unbinding following a quantum quench}

\author{L.~Mathey$^1$ and  A.~Polkovnikov$^2$}

\affiliation{$^1$Joint Quantum Institute, National Institute of Standards and Technology and University of Maryland,
 Gaithersburg, MD 20899\\
$^2$Department of Physics, Boston University, 590 Commonwealth Ave., Boston, MA 02215}

\date{\today}

\begin{abstract}
  We study the dynamics of the relative phase of a bilayer of
  two-dimensional superfluids after the two superfluids have been
  decoupled, using truncated Wigner approximation. On short time
  scales the relative phase shows ``light cone'' like thermalization
  and creates a metastable superfluid state, which can be
  supercritical. On longer time scales this state relaxes to a
  disordered state due to dynamical vortex unbinding.  This scenario
  of dynamically suppressed vortex proliferation constitutes a {\it
    reverse-Kibble-Zurek effect}. We observe dynamics of creation of
  vortex anti-vortex pairs and their consequent motion. Our
  predictions can be directly measured in interference
  experiments~\cite{zoran}.
\end{abstract}


\maketitle

The understanding of order is one of the main objectives of many-body
theory. Apart from topological order and exotic order~\cite{wen}, most
phases can be characterized through the long-range behavior of their
correlation functions.  For example, a system of a single species of
bosons in two spatial dimensions at finite temperature shows two
different regimes: at low temperatures the correlation function of the
boson operator $b(x)$ decays as a power-law $G(x) \equiv \lav
b(x)^\dagger b(0)\rav \sim |x|^{-\tau/4}$, with a scaling exponent
$\tau$, at higher temperatures it decays exponentially $\lav
b(x)^\dagger b(0)\rav \sim \exp(-|x|/x_0)$, with some coherence length
$x_0$.  The transition between these two regimes, occurring at the
critical exponent $\tau_c=1$, is the famous
Berezinsky-Kosterlitz-Thouless (BKT) transition~\cite{BKT}.  It has
been recently observed in ultra-cold atom systems~\cite{zoran, clade}.

The power-law scaling of the quasi-superfluid state is due to
thermally excited phonons. The gapless spectrum and the low
dimension of the system lead to a fluctuating phase at all finite
temperatures.  If these were the only excitations of importance, the
system would show power-law scaling at any temperature with the
exponent $\tau$ being proportional to the temperature $T$ (see
e.g. \cite{cl}). However, vortices can drive the system to a
disordered state, in which the correlation function decays
exponentially. In the quasi-superfluid phase vortex-antivortex (V-AV)
pairs lead only to a small renormalization of the scaling
exponent. These V-AV pairs are 'confined', i.e., they do not
separate spatially significantly.  However, above the transition
temperature, due to entropic effects and due to screening, these
excitations become 'deconfined', i.e., vortices and antivortices
can separate spatially.

\begin{figure}
\includegraphics[width=6.2cm]{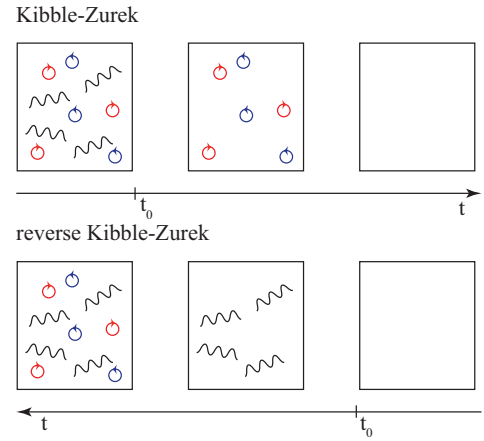}
\caption{Illustration of the Kibble-Zurek mechanism, which describes
 ramping across a phase transition from the disordered phase, and
 its counterpart, the reverse-Kibble-Zurek mechanism. The defining feature of the latter is the dynamical suppression of vortex unbinding, which happens long after phonons reach steady quasi-equilibrium state.
}\label{KZsketch}
\end{figure}

In this paper we study the dynamics of such a system, following a fast
ramp across the transition, coming from the ordered side.  We imagine
two 2D superfluids that are coupled by a tunneling energy $J_\perp$,
which leads to a phase-locked superfluid (SF) state, in which the
fluctuations of the relative phase are suppressed, while the total
phase correlations show algebraic scaling~\cite{mathey07c}.  The
critical temperature of this state typically lies well above the BKT
temperature $T_c$ of the uncoupled system.  We ask the question how
the system transitions to its new equilibrium state after the two SFs
have been decoupled by turning off $J_\perp$. (The case of 1D SFs has
been studied in Refs.~\cite{joerg1, burkov, rafi, mazets}.) We find
that on short time scales vortices are not important and the system
shows 'light-cone'-like behavior in accord with prediction of
Ref.~\cite{cardy} for 1D systems. Qualitatively this behavior comes
from the fact that at distances larger than the product of a
characteristic velocity and the time the correlation functions are not
causally connected and thus decay in time but do not depend on the
distance. Conversely at distances smaller than this product the
correlation functions freeze in time and depend only on distance
between the two points.


The state that emerges after this vortex free evolution is a
metastable SF state. The correlation function shows algebraic
scaling with some 
 exponent that can be related to
the initial temperature and 
the coupling energy of
the initial state.  However, the exponent of that state can be
supercritical, that is, the correlation function can fall off faster
than $|x|^{-1/4}$. This state can not exist in equilibrium and can be
thought of as a 'superheated' superfluid.


On longer time scales this metastable state will relax to thermal
equilibrium phase. If the effective temperature of the metastable
superfluid is small then vortices do not unbind and thus the thermal
equilibrium remains superfluid. If the metastable state is overheated
then the relaxation to equilibrium is accomplished by vortex unbinding
and the system becomes thermal Bose gas.

In Fig. \ref{KZsketch} we illustrate this process, and contrast it to
the Kibble-Zurek (KZ) mechanism \cite{kibble, zurek}. The latter
refers to the case of ramping across a phase transition, from the
disordered side. For a 2D superfluid that would be the thermal Bose
gas phase, in which vortices are deconfined. If the system is then
ramped across the transition with a fast quench, some of these
vortices can survive on a very long time scale, their recombination
with anti-vortices is suppressed
(upper part of Fig.~\ref{KZsketch}).  The case that we consider here is
ramping across the transition from the ordered side. The coupling
$J_\perp$ between the two layers suppresses both phonons and
vortices. When the system is decoupled phonons propagate very fast
creating a metastable quasi-equilibrium state
 (second panel of the lower part of Fig. \ref{KZsketch}).  Only much
later equilibration of vortices occur, leading to thermal
equilibrium. This clear separation of time scales allows us to refer
to this process as the reverse-Kibble-Zurek (rKZ) mechanism.

We emphasize that despite the BKT transition is driven by thermal
fluctuations, the mechanism of vortex or phonon creation in the
process we consider can be triggered by initial quantum
fluctuations. Indeed when the superfluids are strongly coupled
together and the initial temperature is small the density (the
momentum conjugate to the phase) has large fluctuations because of the
zero point motion. The heating mechanism of this system can be thought
of as enhancement of this zero point motion following the quench. This
enhancement of fluctuations can not be captured within conventional
mean-field approaches.

To model
 the relative phase dynamics
 we consider an XY-model~\cite{mathey07c} and
 include a hopping term
\bea
H & = &  \Omega_0\Big(-\sum_{<ij>}\frac{\kappa}{\pi}\cos(\phi_i-\phi_j)
 + \frac{\pi}{2 \kappa} \sum_{i} n_{i}^2\nonumber\\
& & - V(t) \sum_{i} \cos(\sqrt{2}\phi_i)\Big),
\label{hamiltonian}
\eea
where $\Omega_0$ is an overall (Josephson) energy scale, $\kappa$
describes the ratio of kinetic and potential energies.  We can
formally replace these parameters by $\Omega_0\kappa/\pi = 2 J n$,
$\pi \Omega_0/\kappa = U$ (so that $\Omega_0=\sqrt{2JnU}$,
$\kappa=\pi\sqrt{2Jn/U}$) and $V(t) = 2 J_\perp(t) n/\Omega_0$, which
gives a coarse-grained representation of the Hubbard model, where the
Bose operators have been written in a phase-density representation. In
the Hubbard model $J$ is the in-plane hopping, $U$ is the interaction
energy, $n$ is the average number of bosons per site, and $J_\perp$ is
an inter-layer hopping. In the continuum limit the lattice size is
approximately given by the healing length in the system.

\begin{figure*}
\includegraphics[width=16cm]{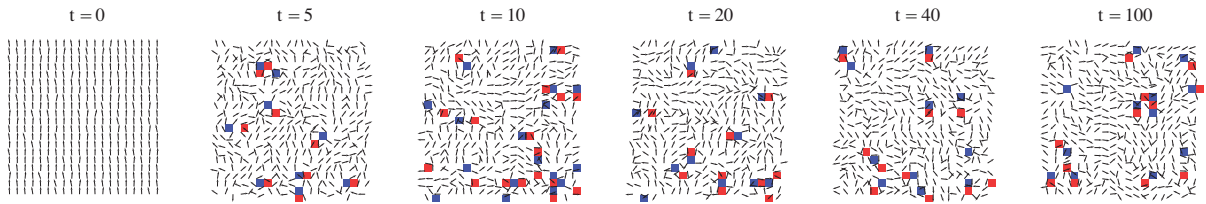}
\caption{\label{run}
Temporal evolution of the relative phase for a single realization of initial conditions. The parameters of the systems are  $V=100$, $\kappa=10$ and $T_0=2$. Vortices are marked red, anti-vortices blue.
}
\end{figure*}

%

%
It is convenient to introduce the rescaled quantities $\tilde{t} =
\Omega_0 t/ \hbar$, $\tilde{\phi} = \sqrt{\frac{\kappa}{\pi}} \phi$,
and $\tilde{n} = \sqrt{\frac{\pi}{\kappa}} n$.  In terms of these, the
equations of motion (EOMs) are
\bea\label{eom}
\frac{d\tilde{\phi}_i}{d\tilde{t}} & = & - \tilde{n}_i\nonumber\\
\frac{d\tilde{n}_i}{d\tilde{t}} & = & -
\frac{\sqrt{2}}{\beta}\sum_{j_i}\sin\Big(\frac{\beta(\tilde{\phi}_{j_i}-
  \tilde{\phi}_i)}{\sqrt{2}}\Big) + V(t) \beta \sin \beta
\tilde{\phi}_i,
\eea
where we defined $\beta = \sqrt{2\pi/\kappa}$. The indices $j_i$
describe the four neighboring sites of site $i$.

We model the relative phase using a numerical implementation of the
truncated Wigner approximation (TWA) (see Refs.~\cite{blakie} for a
review): The expectation of any quantity at some time $t>0$ can be
determined by sampling over a Wigner distribution at time $t=0$, and
solving the classical equations of motion from $0$ to $t$. This
approximation is accurate either at short times, or with nearly
harmonic systems~\cite{ap_twa}. In our case initial ``light-cone''
stage of dynamics is well described within the quadratic Bogoliubov
theory where TWA is exact. At longer times, when nonlinear dynamics
takes over, we expect TWA to remain applicable because by that time
relevant momentum modes become highly occupied justifying the validity
of the approach~\cite{blakie}. The advantage of this method is that
each run strongly resembles a single realization of experiment,
especially for a positively defined initial Wigner function. Thus
to a very good precision one can mimic actual experiments. In
Fig.~\ref{run} we plot time evolution of the relative phase for a
single realization on initial conditions. One can observe V-AV pair
formation at short times and their consequent unbinding.

\begin{figure}
\includegraphics[width=8.4cm]{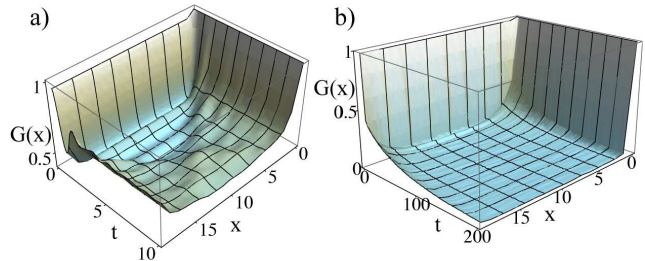}
\caption{\label{GF}
Plot of short (left) and long (right) time behavior of
the correlation function as a function of time and space. The parameters a
$T_0=3$, for $\kappa=10$ and $V=20$ for the left graph and $T_0=1$, $\kappa = 8$, and $V=80$ for the right graph. At short times the dynamics separates into instantaneous, damped oscillatory behavior, and a 'light cone' like pulse forming metastable quasi-superfluid state. At longer time scales the correlation function
  shows exponential decay due to dynamical  vortex unbinding.
}
\end{figure}
We can calculate the Wigner distribution at $t=0$,  assuming 
that $J_\perp$ is larger than the other energy scales.  In this limit
the phase fluctuations can be described within the Bogoliubov
approximation, so the system reduces to a sum of oscillators.  The
Fourier modes $\tilde{\phi}_q$ and $\tilde{n}_q$ at $t=0$ are
distributed according to (see Ref.~\cite{adiabatic})
\bea
W & \sim & \exp\Big(-\frac{\tilde{\phi}_q^2}{2 \sigma_q^2 r_q}
 -\frac{2 \sigma_q^2\tilde{n}_q^2}{r_q}\Big)
 \label{wig}
\eea
with $\sigma = 1/\sqrt{2\omega_q}$, $r_q = 1/\tanh(\omega_q/2T_0)$,
and $\omega_q = \sqrt{4 \sin(q_x/2)^2 + 4 \sin(q_y/2)^2 + V\beta^2}$,
$T_0$ being initial temperature.
%
%
We use this method to extract the equal time correlation function:
\beq
G(x,t)=\langle \exp[i\sqrt{2}\phi_j(t)-i\sqrt{2}\phi_{j+x}(t)]\rangle,
\eeq
where $x$ is an integer separation between the points and $t$ is the
time after decoupling (see Fig.~\ref{GF}). Because we are using
periodic boundary conditions $G(x,t)$ depends only on the separation
between the points $x$. Note that this correlation function (or rather
$\int_0^x dx' G(x',t)$) can be directly measured in interference
experiments~\cite{interference, zoran, joerg1}. We indeed see very
clear emergence of the light cone thermalization: At separations
larger than $2vt$, where $v$ is characteristic phonon velocity,
$G(x,t)$ is almost $x$ independent - it uniformly decreases in
time. Once $2vt>x$ the correlations freeze in time and depend only on
$x$. We find that the state that emerges within the light cone shows
algebraic scaling, and therefore can be referred to as a
quasi-superfluid. At much longer times scale the correlation function
clearly relaxes to the exponential equilibrium shape due to vortex
unbinding.

\begin{figure}
\includegraphics[width=7.6cm]{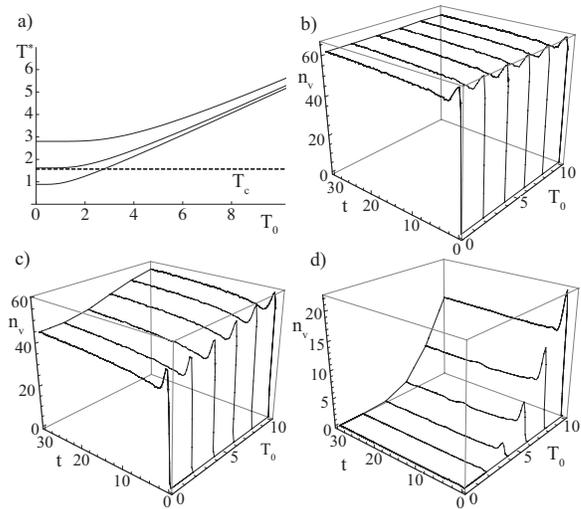}
\caption{\label{quenchsim} a) $T^\star$, as given in
  Eq. \ref{Tstarapp}, for $\kappa=1,3,10$, from top to bottom, and for
  $V=20$. The line $T_c=\pi/2$ is the BKT 
  temperature of the uncoupled SFs.  b) -- d) Simulations for these values of $\kappa$ and
  $V$.  }
\end{figure}

Due to the relation between the power-law exponent and temperature in
equilibrium, we can associate an effective temperature $T^\star$ with
the state inside the light cone.  We can estimate this effective
temperature by considering the linearized dynamics, which neglects
vortices.  Then Eqs.~(\ref{eom}) have the solution:
\bea
{\tilde\phi}_k(t) & = & {\tilde\phi}_k(0) \cos \omega_k {t} - {\tilde n}_k(0)/\omega_k \sin \omega_k {t}\\
{\tilde n}_k (t) & = & {\tilde\phi}_k(0) \omega_k \sin \omega_k {t} + {\tilde n}_k(0) \cos \omega_k {t}.
\eea
 We then consider the quadrature
\bea
\lav\phi_k^2({t}) \rav
& = &  \frac{r_{k,0}}{2\omega_{k,0}} \cos^2(\omega_k {t})
 + \frac{r_{k,0} \omega_{k,0}}{2\omega_k^2}\sin^2(\omega_k {t}).
\eea
where $r_{k,0}$ and $\omega_{k,0}$ correspond to the quantities at
$t=0$, that is to $V(0) =V$, and $\omega_k$ to the dispersion with
$V=0$.
%
The long-time limit $\lav\phi_k^2({t} \rightarrow \infty)\rav$ is
$r_{k,0}/4\omega_{k,0} + r_{k,0} \omega_{k,0}/4\omega_k^2$.
 We equate this formally to a thermal ensemble of momentum-dependent
 'temperature'  $T^\star_k$, i.e.
 $\lav\phi_k^2({t} \rightarrow \infty)\rav =  r_k^\star/\omega_k$,
%
%
where $r_k^\star = 1/\tanh(\omega_k/2T^\star_k)$.
 We solve for $T_k^\star$, and find that
%
%
 for large $V\beta^2$ it simplifies to a single value, independent of
 $k$:
\bea\label{Tstarapp}
T^\star & = & \frac{\sqrt{V \beta^2}}{4 \tanh(\sqrt{V\beta^2}/2T_0)}.
\eea
For small $T_0$ we have $T^\star \approx \frac{\sqrt{V \beta^2}}{4}$
($T^\star=2J_\perp/J$ in terms of original Hubbard parameters), that
is, the temperature is fully determined by the initial coupling
energy. The coupling energy between the two layers is transferred into
the in-plane kinetic energy.  We note that if $J_\perp$ exceeds the
chemical potential $U n$ the quantum rotor model becomes inadequate
and then $T^\star$ saturates at the value $T^\star\sim Un/J$. For
large $T_0$ we have $T^\star \approx \frac{T_0}{2}$.  This is a reflection
of the doubling of the degrees of freedom when two layers are
uncoupled.
In Fig. \ref{quenchsim} a) we plot $T^\star$ for $V=20$ and for
different value of $\kappa$.  For $\kappa=1$ corresponding to large
$J_{\perp}$, $T^\star$ is always above the critical temperature $T_c
=\pi/2$, for $\kappa=10$, it crosses it as initial temperature
increases.  We therefore expect to see very little vortex formation
for small initial temperatures for $\kappa=10$, and many vortices for
all temperatures for $\kappa=1$. This is indeed the case as we show in
Fig. \ref{quenchsim}, which plots the number of vortices $n_v$ versus
time and initial temperature for three different values of
$\kappa$. This number is obtained by counting the plaquettes with a
phase winding of $2\pi$, and then by averaging over many runs.
\begin{figure}
\includegraphics[width=5.2cm]{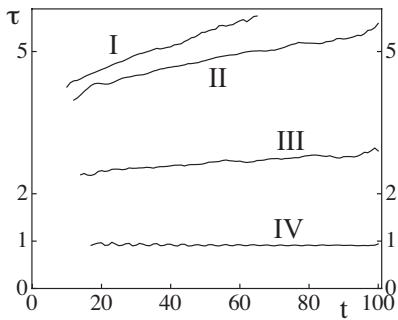}
\caption{\label{fit} Exponent $\tau$ extracted from fitting   the correlation function $G(x)$ to the algebraic form, for different initial couplings. We use $T_0=1$ and $\kappa = 8$, and $V=80, 70, 50$, and $20$, corresponding to curves I to IV.  Except for the curve I corresponding to Fig.~\ref{GF}-b) the correlation function can be fitted with an algebraic function throughout the time interval shown here.}
\end{figure}
%




To quantify the crossover from the supercritical superfluid state to
the normal phase we fit the correlation function $G(x,t)$ at different
times numerically using algebraic ($G(x,t)=c(L/\pi|\sin(\pi
x/L)|)^{-\tau/4}$) and an exponential ($G(x,t)=c \exp(-|\sin(\pi
x/L)|/a)$) functions.  In equilibrium the algebraic exponent $\tau$
would be the relative temperature $T/T_c$. Any value above $1$ is
therefore supercritical.  We use these two fitting functions in four
examples with $T_0=1$ and $\kappa=8$, but with different initial
couplings $V=80, 70, 60, 20$. In Fig. \ref{fit} we show the exponent
$\tau$ as a function of time for these cases.  In all cases $G(x,t)$
develops algebraic scaling after the light-cone dynamics. Note that
for large initial couplings (I-III) the emerging scaling exponent
$\tau$ is well above the critical exponent. At longer times the
exponent slowly increases in time. During this process, the
correlation function can still be well fitted with an algebraic
function. Eventually the correlation function develops exponential
scaling.  This regime is reached for $V=80$ (I) within the time
interval shown in Fig. \ref{fit}, signalling that the thermal Bose gas
phase has been reached due to vortex unbinding.  For $V=70$ (II) and
$V=50$ (III) the time scale of vortex unbinding is longer then the
time interval shown.  For $V=20$ (IV) the system remains in a
quasi-superfluid state. We conclude from these examples that the
supercritical superfluid has a long life-time and the exponent $\tau$
can significantly exceed the maximum equilibrium value. Such
metastable states should be thus experimentally feasible.

In conclusion, we have studied the dynamics of the relative phase of a
bilayer of superfluids in 2D, after the hopping between them has been
turned off rapidly.  We find that on short time scales the dynamics of
the correlation function shows a ``light-cone''-like behavior. The
system then develops an algebraic phase that can be stable or
metastable. The latter can be thought of as a superheated
superfluid. On long time scales this metastable superfluid relaxes to
a disordered state via creating of vortex-antivortex pairs and their
consequent unbinding.  

Our predictions can be directly probed in
experiment. The behavior of the relative of phase of two
superfluids can be studied by interference experiments \cite{interference},
 so both the supercritical state and the algebraic-to-exponential
 evolution can be tested. Vortex unbinding can also be demonstrated
 by direct observation of free vortices \cite{zoran}.
 The time scale of the light-cone dynamics can be estimated
 as $t_{LC}\sim L/v$, where the system size $L\sim 10^{-4}$m and
 the phonon velocity $v\sim 10^{-3}$m/s leads to $t_{LC}\sim 0.1$s.
 The rate of vortex unbinding is exponentially suppressed
 compared to the BKT energy scale~\cite{longpaper}.

 \acknowledgements The authors acknowledge useful discussions with
 A.H. Castro Neto and E. Altman.  A.P. was supported by AFOSR YIP and
 Sloan Foundation.  L.M. acknowledges support from NRC/NIST,
  NSF Physics Frontier Grant PHY-0822671 and 
 Boston University visitor's program.

%
%

\def\etal{\textit{et al.}}

\end{document}